\documentclass[times,amssymb,usenatbib,useAMS,twocolumns]{mn2e}
\usepackage[utf8]{inputenc}
\usepackage{amsmath}
\usepackage{amsfonts}
\usepackage{graphicx}
\usepackage{ragged2e}
\justifying
\title[Dehydrogenated PAH anions]{Interstellar dehydrogenated PAH anions: vibrational spectra}
\author[Buragohain et al.]{Mridusmita Buragohain$^{1}$\mbox{*}, Amit Pathak$^{2}$, Peter Sarre$^3$, Nand Kishor Gour $^{4}$\\
$^{1}$Department of Physics, Tezpur University, Tezpur 784028, India (ms.mridusmita@gmail.com, \mbox{*}Corresponding author)\\
$^{2}$Department of Physics, Banaras Hindu University, Varanasi 221 005, India (amitpah@gmail.com)\\
$^{3}$School of Chemistry, The University of Nottingham, University Park, Nottingham, NG7~2RD, United Kingdom\\
$^{4}$Department of Chemical Sciences, Tezpur University, Tezpur 784\,028, India \\}

\begin{document}
\date{}
\maketitle
\begin{abstract}
Interstellar Polycyclic Aromatic Hydrocarbon (PAH) molecules exist in diverse forms depending on the local physical environment.
Formation of ionized PAHs (anions and cations) is favourable in the extreme conditions of the ISM. Besides in their pure form, PAHs
are also likely to exist in substituted forms; for example, PAHs with functional groups, dehydrogenated PAHs etc. 
A dehydrogenated PAH molecule might subsequently form fullerenes in the ISM as a result of ongoing chemical processes. 
This work presents a Density Functional Theory (DFT) calculation on dehydrogenated PAH anions to explore
the infrared emission spectra of these molecules and discuss any possible contribution towards observed IR features
in the ISM. The results suggest that dehydrogenated PAH anions 
might be significantly contributing to the 3.3~$\mu \rm m$ region. Spectroscopic features unique to dehydrogenated PAH anions are highlighted
that may be used for their possible identification in the ISM.
A comparison has also been made to see the size effect on spectra of these PAHs.
\end{abstract}


\section{Introduction}
Mid-infrared emission lines popularly known as `Aromatic Infrared Bands' (AIBs) are broad emission features widely distributed
in the Interstellar Medium (ISM) and observed at 3.3, 6.2, 7.7, 8.6, 11.2, 12.7 and 16.4~$\mu \rm m$ towards diverse 
astrophysical objects \citep[]{Tielens08}. The regions where the AIBs are observed
include H~\textsc{ii} regions, reflection nebulae, planetary nebulae, 
photodissociation regions, AGB objects, active star forming regions, young stellar objects, the diffuse interstellar medium, external galaxies etc. 
\citep[]{Onaka96, Verstraete96, Hony01, Peeters02, Acke04, Sakon04, Regan04, Brandl06, Armus07}. Lately, 
AIBs have emerged as an acclaimed area of research due to numerous detections in the ISM and the ambiguity associated in 
the identification of their carriers. It was first proposed by \citet[]{Duley81} that these features arise from the thermal emission by small 
carbon grains with functional groups. This hypothesis was later revised by \citet[]{Leger84} and \citet[]{Allamandola85} to free, gas-phase Polycyclic Aromatic 
Hydrocarbon (PAH) molecules, giving rise to these features in the mid-IR as a result of vibrational relaxation 
on absorption of background UV photons. 
The PAH-AIB hypothesis is now very widely accepted, though uncertainty lies in the identification of the exact 
form and functional groups attached to the PAH molecule. An increasing
number of observations point to a wide PAH family range rather than a single conventional 
form which may explain the complete set of AIBs \citep[]{Snow98, Peeters04, Hudgins05, 
Simon11, Onaka14, Mridu15}. A recent study by \citet[]{Yang17a} 
has proposed interstellar PAH molecules to be a mixture of aromatic/aliphatic components. 

PAH molecules have also been suggested to be potential candidates 
for producing absorption features in the optical region known as `Diffuse Interstellar Bands (DIBs)'
\citep[]{Sarre06, Cox06, Pathak08, Salama11} and may be contributing significantly to other important interstellar phenomena, 
e.g., heating of the ISM and charge balance inside molecular clouds \citep[]{Lepp88, Verstraete90, Bakes94, Bakes01a, Bakes01b}.

The AIBs vary in terms of peak position, band width and intensity as a function of the local physical condition of the observed sources  
\citep[]{Hony01, Peeters02, Tielens08}. An overall correlation among the bands is consistent across the source environment.
It is concluded that neutral, cationic as well as anionic PAHs contribute to the emission of some infrared bands depending
on the interstellar conditions \citep[]{Bauschlicher08, Bauschlicher09}. 
Emission at 6.2, 7.7 and 8.6~$\mu \rm m$ are mostly contributed by PAH cations whereas the 3.3 and 11.2~$\mu \rm m$ bands preferentially arise
from neutral PAHs in the ISM \citep[]{Peeters02, Tielens08, Schmidt09}. 
Cationic PAHs show a comparatively weaker 3.3~$\mu \rm m$ compared to neutrals.
PAH cations are abundant in intense
UV irradiated fields where PAHs are predominantly affected by photoionization \citep[]{Omont86}. In contrast to this, a comparatively less intense
UV irradiated field can leave a PAH negatively charged, or an anion formed by addition of an electron \citep[]{Omont86}. 
Observational evidence for negatively charged molecules in the dense molecular cloud TMC-1 adds strength 
to the probable existence of astrophysical PAH anions \citep[]{McCarthy06}. Several theoretical 
and experimental studies \citep[]{Lepp88-a, Szczepanski95,
Bakes01a, Bakes01b, Ruitercamp02, Wang05, Malloci05, Malloci07} discuss the 
role of astrophysical PAH anions heating the ISM, free electron density, 
recombination processes, etc.
\citet[]{Herbst81} and \citet[]{Herbst08} showed that negative ions are formed through radiative 
attachment of electrons to neutral species in dense clouds. In a similar way, PAHs can significantly bear a significant fraction of 
the negative charge which then become the principal
carriers of the negative charge in the cold interstellar medium. When an electron is attached to a PAH molecule, 
the resultant can be a transient negative ion [PAH$_{n}^-$]*
which further can decay into a stable full anion PAH$_{n}^-$ or to a dehydrogenated PAH anion (PAH$_{n-1}^-$) through subsequent 
loss of an H atom \citep[]{Garcia13}. Besides, dehydrogenated PAH molecules have larger electron affinities 
compared to the parent hydrocarbons which indicates a significant presence of
dehydrogenated PAH anions in environments like diffuse or circumstellar media \citep[]{Hammonds11}. 
\citet[]{Tobita92} showed the formation of dehydrogenated PAH anion from a PAH molecule and
reported that dehydrogenated PAH anions are the only possible fragment anions formed. Thus, any astrophysical model that
takes into account interstellar PAH anions should also include dehydrogenated forms. The dehydrogenation state of PAH molecules in the ISM
is further endorsed by the recent detection of fullerenes as carriers for two observed DIBs \citep[]{Campbell15, Ehrenfreund15}. It is
proposed that cosmic fullerenes may be formed from interstellar PAHs in a top-down or a bottom-up process \citep[]{Dunk12, Bernes12}. 
In either of the two formation processes initiated by PAHs, dehydrogenated PAHs are an intermediate form \citep[]{Mackie15} 
which may subsequently be altered into other forms; for example, a dehydrogenated PAH anion through radiative attachment of electrons in dense cloud.

This work reports a Density Functional Theory (DFT) of PAH anions (PAH$_{n}^-$) along with dehydrogenated forms (PAH$_{n-1}^-$) to 
study their possible spectral characteristics in the infrared in relation to the AIBs. 
PAH anions are astrophysically important as these have more stable configuration (in terms of energy) 
compared to cations and neutrals under interstellar conditions. The size effect on the intensity of 
vibrational transitions of these PAHs is also studied in this report. 

\section{Theoretical Approach}
Compared to laboratory experiments,
theoretical quantum chemical calculations have appeared to be more feasible to study the chemical and
physical properties of PAH molecules.
Density Functional Theory (DFT) is an appropriate quantum chemical method that has been used rigorously
to calculate the vibrational properties of interstellar molecules in order to compare with the observed AIBs 
\citep[]{Langhoff96, Hudgins04a, Pathak05, Pathak06, Pathak07, Pauzat11, Pauzat11-a, Candian14}.
DFT can be applied on a wide range of PAH molecules of varying size, charge and functional groups including
molecules that are not easily synthesized in the laboratory. An appropriate combination of
basis set and functional is chosen to compromise between accuracy of the results and computational demand \citep[]{Yang17a}. 
For the present study, we are performing 
DFT combined with B3LYP/6-311G** basis on PAH molecules to calculate the optimized structure of the molecules which is
further used to obtain the harmonic frequencies and intensities of vibrational modes at the same level
of theory. Vibrational modes in a PAH molecule are of different types, for example : C$-$H$\rm_{oop}$ (out of plane), C$-$H$\rm_{in~plane}$,
C$-$C$\rm_{stretch}$ and C$-$H$\rm_{stretch}$. 
The frequencies obtained from the results of quantum chemical
calculation are overestimated as compared to the experimental data. 
In order to match with the experimental frequencies, computed frequencies are scaled down with mode-dependent 
scaling factors.
A scaling factor of 0.974 for C$-$H$\rm_{oop}$,
0.972 for C$-$H$\rm_{in~plane}$ and C$-$C$\rm_{stretch}$ and 0.965 for C$-$H$\rm_{stretch}$ is used \citep[]{Mridu15, Mridu16}. 
These scaling factors are calculated by comparing experimental and theoretical frequencies.
When the relative strengths of the modes obtained from theoretical calculation are compared with those obtained from experiments,
the intensity of C$-$H stretch is found to be much larger as compared to the other modes and an intensity scaling is required.
Considering that the second order M\o{}ller-Plesset (MP2) perturbation theory with large basis set (for example, MP2/6-311+G(3df, 3pd)) gives more accurate
oscillator strengths compared to B3LYP DFT, \citet[]{Yang17b} have derived a relation for MP2/6-311+G(3df, 3pd) and B3LYP/6-311+G** 
level of theories to scale the
intensities of C$-$H$\rm_{stretch}$ modes near $\sim$~3~$\mu \rm m$ region. There are two important features in this region;
3.3~$\mu \rm m$ and 3.4~$\mu \rm m$ due to the stretching of aromatic C$-$H and aliphatic C$-$H bonds respectively. 
By using, A$_i$~{$\approx$}~0.6372 A$_j$, where A$_i$ and A$_j$ are the intensities of C$-$H$\rm_{stretch}$ modes computed at the MP2/6-311+G(3df, 3pd) and
B3LYP/6-311+G** level, respectively, we can achieve good accuracy by computing at an inexpensive level \citep[]{Yang17b}. 
Relative intensities (Int$\rm_{rel}$\footnote{Int$\rm_{rel}$
=$\frac{\rm{absolute~intensity}}{\rm{maximum~absolute~intensity}}$}) are obtained by taking the ratio of all intensities to the maximum intensity 
for normalization. The computed position (in wavelength) and relative intensity (Int$\rm_{rel}$) of the lines are 
used to plot a Gaussian profile with a FWHM of 30 cm$^{-1}$. The profile width is typical for PAHs emitting in an interstellar environment 
and depends on vibrational energy redistribution of the molecule \citep[]{Allamandola89}. 
In some cases, a few overlapping bands in a particular region (mostly in the 3~$\mu \rm m$ region) 
might add up producing band intensities to exceed unity
in the resulting spectra \citep[]{Mridu15, Mridu16}. 
It is important to note here that for some species, several (at least 2) spin-multiplicities are possible 
(for C$_{24}$H$_{11}^{-1}$ for instance, singlet and triplet spin states are a priori possible). However, as a preliminary study,
we have limited our calculation to molecules with the lowest possible spin-multiplicity. 
This is because the sample molecules that we have considered exhibit the lowest energy with the lowest possible spin-multiplicity 
corresponding to the most stable configuration to endure in the interstellar domain \citep[]{Bausch13, Theis15, Fortenberry16}. 
For example, for C$_{24}$H$_{11}^{-1}$, lowest lying state is a singlet state.
 We have used QChem (quantum chemistry package) to perform our calculation \citep[]{YSaho15}.
 \section*{Results and Discussion}
\begin{figure*}
\vspace*{-2em}
\centering
\includegraphics[width=12cm, height=14cm]{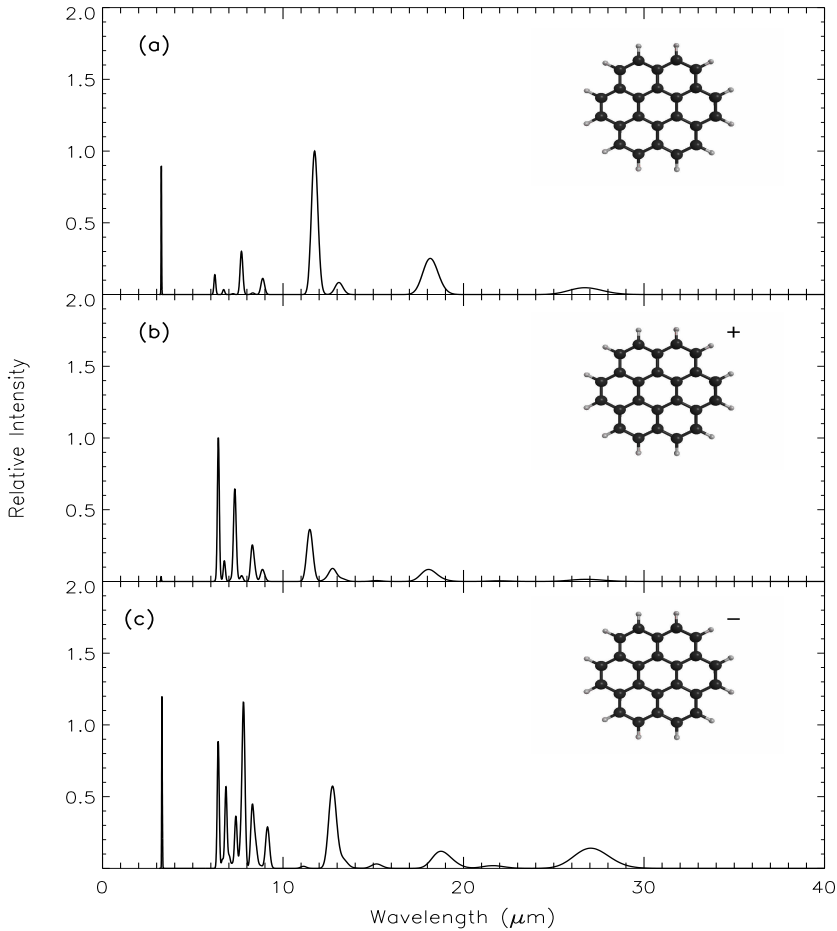}
\caption{Theoretical spectra of (a) neutral coronene (C$_{24}$H$_{12}$), 
(b) cationic coronene (C$_{24}$H$_{12}^+$), (c) anionic coronene (C$_{24}$H$_{12}^-$)}
\label{fig1} 
\end{figure*}

Coronene (C$_{24}$H$_{12}$) is a mid-sized compact PAH molecule and is a good representative of the interstellar PAH family. 
For its stable configuration due to the delocalization of the electrons 
and its compactness, coronene is considered as a good choice for our study. 
We compare the IR spectra of neutral coronene (C$_{24}$H$_{12}$) with its cationic (C$_{24}$H$_{12}^+$) and anionic (C$_{24}$H$_{12}^-$) forms
in Figure~\ref{fig1}. The features are characteristic of various vibrational modes present in the respective molecules.
Table~\ref{tab1} presents the wavelengths corresponding to different modes of coronene in its neutral and ionized forms.
While the wavelengths of these bands are less affected by ionization (positive or negative), 
a distinct variation in the intensity is produced on ionization of the molecule. This is 
similar to the results discussed in \citet[]{DeFrees93, Langhoff96, Pauzat97}.
Intensities of bands in the 6$-$10~$\mu \rm m$ regions which are modest
in C$_{24}$H$_{12}$ become significantly more prominent in C$_{24}$H$_{12}^+$ and C$_{24}$H$_{12}^-$ (Figure~\ref{fig1}b \& Figure~\ref{fig1}c. 
The features in the 6$-$10~$\mu \rm m$ region are characteristic of 
the C$-$C stretch and C$-$H in$-$plane bending vibrations of the PAH molecules. 
The 3.3~$\mu \rm m$ feature due to the C$-$H stretch mode is intense for 
C$_{24}$H$_{12}$ and C$_{24}$H$_{12}^-$ and weak for C$_{24}$H$_{12}^+$.
Intensities of all the 3.3~$\mu \rm m$ features are scaled here (scaling factor $\sim$~0.6372) and the relative intensity of the 3.3~$\mu \rm m$ feature in cation is
more consistent with the astronomical 3.3~$\mu \rm m$ band.
Previous laboratory and theoretical studies \citep[and references therein]{Tielens08} 
reveal that the astronomical 3.3~$\mu \rm m$ feature is likely 
to be dominated by the C$-$H stretch modes in neutral PAH molecules, whereas 5$-$10~$\mu \rm m$ region arises due to the
C$-$C stretch and C$-$H in$-$plane bending modes in cationic PAH molecules. PAH anions also contribute to the 5$-$10~$\mu \rm m$ region 
and to the 3.3~$\mu \rm m$ feature which indicates them to
be an important constituent of the ISM. Importantly anionic PAHs show a comparatively 
strong 3.3~$\mu \rm m$ feature.

The band positions and intensities beyond 15~$\mu \rm m$ are affected to a lesser extent by ionization, both positive and negative.
For example: a feature near 18~$\mu \rm m$ in neutral and cationic C$_{24}$H$_{12}$ which is a combination mode of 
C$-$C$-$C oop and C$-$H oop bending is shifted to 18.7~$\mu \rm m$ for anionic C$_{24}$H$_{12}$. Similarly, a variation in
intensity is also observed, particularly a feature near 28~$\mu \rm m$, 
which is a C$-$C$-$C in$-$plane vibration which distinctly appears in the spectrum of 
C$_{24}$H$_{12}^-$ unlike its neutral and cationic counterparts. These features in the 
longer wavelength side are not much studied experimentally.
\begin{table*}
\centering
  \begin{minipage}{150mm}
\caption{Theoretical spectral data for neutral, cationic and anionic coronene}
\label{tab1}
\resizebox{\textwidth}{!}{%
\newcommand{\mc}[3]{\multicolumn{#1}{#2}{#3}}
\begin{tabular}[c]{c|c|c|c|c|ccc|}
\hline
\hline
PAH & Frequency & Wavelength & Relative& \mc{4}{c}{Mode} \\
 & ($\rm cm^{-1}$) & (~$\mu \rm m$) & Intensity & \mc{4}{c}{} \\ \hline
 & 3068.32 & 3.3 & 0.43 & \mc{4}{c}{$\rm{C-H}$ stretch} \\
 & 3067.74 & 3.3 & 0.44 & \mc{4}{c}{$\rm{C-H}$ stretch} \\
  & 1607.23 & 6.2 & 0.07 & \mc{4}{c}{$\rm{C-C}$ stretch} \\
Neutral & 1606.67 & 6.2 & 0.07 & \mc{4}{c}{$\rm{C-C}$ stretch} \\
coronene  & 1300.34 & 7.7 & 0.15 & \mc{4}{c}{$\rm{C-C}$ stretch + $\rm{C-H}$ in$-$plane} \\
  & 1297.93 & 7.7 & 0.15 & \mc{4}{c}{$\rm{C-C}$ stretch + $\rm{C-H}$ in$-$plane} \\
   & 1127.93 & 8.9 & 0.06 & \mc{4}{c}{$\rm{C-H}$ in$-$plane} \\
   & 851.15 & 11.7 & 1 & \mc{4}{c}{$\rm{C-H}$ oop bending} \\
   & 550.9 & 18.2 & 0.25 & \mc{4}{c}{$\rm{C-C-C}$ oop bending + $\rm{C-H}$ oop bending} \\
 \hline
 & 1558.29 & 6.4 & 1 & \mc{4}{c}{$\rm{C-C}$ stretch} \\
 & 1482.10 & 6.7 & 0.12 & \mc{4}{c}{$\rm{C-C}$ stretch + $\rm{C-H}$ in$-$plane} \\
 & 1377.30 & 7.3 & 0.1 & \mc{4}{c}{$\rm{C-C}$ stretch + $\rm{C-H}$ in$-$plane} \\
 & 1363.26 & 7.3 & 0.58 & \mc{4}{c}{$\rm{C-C}$ stretch + $\rm{C-H}$ in$-$plane} \\
  Coronene & 1206.86 & 8.3 & 0.18 & \mc{4}{c}{$\rm{C-C}$ stretch + $\rm{C-H}$ in$-$plane} \\
  cation & 1206.47 & 8.3 & 0.05 & \mc{4}{c}{$\rm{C-C}$ stretch + $\rm{C-H}$ in$-$plane} \\
  & 1187.09 & 8.4 & 0.07 & \mc{4}{c}{$\rm{C-H}$ in$-$plane} \\
  & 1130.89 & 8.8 & 0.07 & \mc{4}{c}{$\rm{C-H}$ in$-$plane} \\
  & 870.9 & 11.5 & 0.36 & \mc{4}{c}{$\rm{C-H}$ oop bending} \\
  & 784.04 & 12.8 & 0.09 & \mc{4}{c}{$\rm{C-C-C}$ in$-$plane} \\
  & 554.58 & 18 & 0.08 & \mc{4}{c}{$\rm{C-C-C}$ oop bending + $\rm{C-H}$ oop bending} \\
\hline
  & 3040.03 & 3.3 & 0.59 & \mc{4}{c}{$\rm{C-H}$ stretch} \\
 & 3038.47 & 3.3 & 0.57 & \mc{4}{c}{$\rm{C-H}$ stretch} \\
 & 3013.34 & 3.3 & 0.28 & \mc{4}{c}{$\rm{C-H}$ stretch} \\
 & 1563.83 & 6.4 & 0.35 & \mc{4}{c}{$\rm{C-C}$ stretch} \\
 & 1558 & 6.4 & 0.56 & \mc{4}{c}{$\rm{C-C}$ stretch} \\
 & 1505.7 & 6.6 & 0.06 & \mc{4}{c}{$\rm{C-C}$ stretch} \\
 & 1473.22 & 6.8 & 0.05 & \mc{4}{c}{$\rm{C-C}$ stretch + $\rm{C-H}$ in$-$plane} \\
  & 1461.75 & 6.8 & 0.54 & \mc{4}{c}{$\rm{C-C}$ stretch +$\rm{C-H}$ in$-$plane} \\
  & 1424.3 & 7 & 0.08 & \mc{4}{c}{$\rm{C-C}$ stretch +$\rm{C-H}$ in$-$plane} \\
  Coronene& 1353.75 & 7.4 & 0.36 & \mc{4}{c}{$\rm{C-C}$ stretch +$\rm{C-H}$ in$-$plane} \\
  anion & 1305.68 & 7.7 & 0.18 & \mc{4}{c}{$\rm{C-C}$ stretch +$\rm{C-H}$ in$-$plane} \\
  & 1286.69 & 7.8 & 0.15 & \mc{4}{c}{$\rm{C-C}$ stretch +$\rm{C-H}$ in$-$plane} \\
  & 1279.13 & 7.8 & 1 & \mc{4}{c}{$\rm{C-C}$ stretch +$\rm{C-H}$ in$-$plane} \\
  & 1206.34 & 8.3 & 0.37 & \mc{4}{c}{$\rm{C-C}$ stretch +$\rm{C-H}$ in$-$plane} \\
  & 1196.79 & 8.4 & 0.08 & \mc{4}{c}{$\rm{C-H}$ in$-$plane} \\
  & 1177.63 & 8.5 & 0.14 & \mc{4}{c}{$\rm{C-H}$ in$-$plane} \\
  & 1093.85 & 9.1 & 0.29 & \mc{4}{c}{$\rm{C-H}$ in$-$plane} \\
  & 785.14 & 12.7 & 0.48 & \mc{4}{c}{$\rm{C-H}$ oop bending} \\
  & 779.28 & 12.8 & 0.07 & \mc{4}{c}{$\rm{C-C-C}$ stretch} \\
  & 747.96 & 13.4 & 0.05 & \mc{4}{c}{$\rm{C-C-C}$ in$-$plane} \\
  & 535.98 & 18.7 & 0.1 & \mc{4}{c}{$\rm{C-C-C}$ oop bending + $\rm{C-H}$ oop bending} \\
\hline
\end{tabular}
}%
\raggedright
Modes with intensity equal to or greater than 0.05 are only listed in the table. \\
oop is out of plane. \\
intensity of 3.3~$\mu \rm m$ band has been scaled. \\
\end{minipage}
\end{table*}

\begin{figure*}
\hspace*{-5em}
\vspace*{-2em}
\includegraphics[width=15cm, height=18cm]{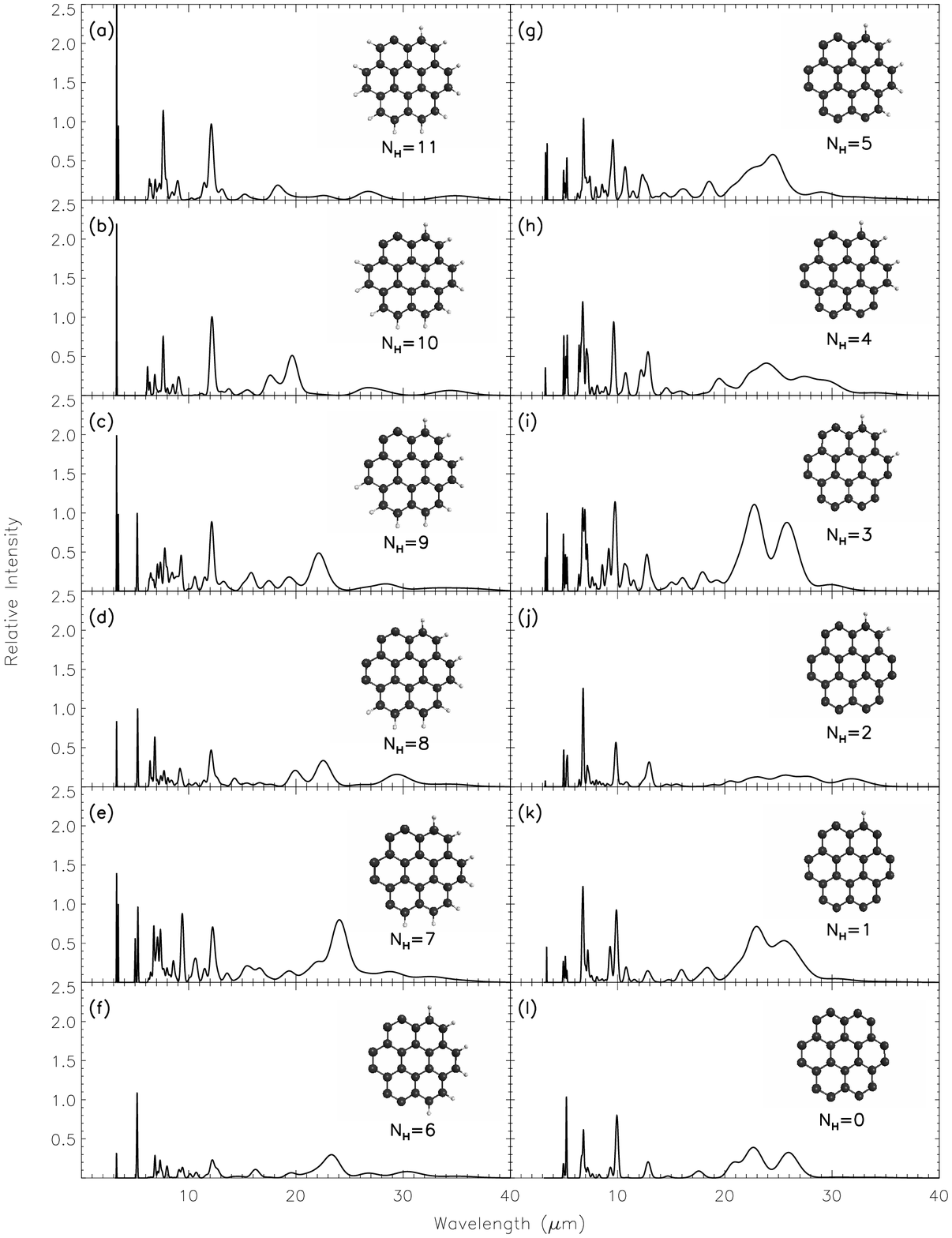}
\caption{Theoretical spectra of (a)-(l) PAH (coronene, C$_{24}$H$_{12}$) anion in dehydrogenated forms}
\label{fig2} 
\vspace*{-0.5em}
\centering{
\raggedright N$\rm_{H}$=number of H atoms in the periphery of a PAH molecule}
\end{figure*}

Dehydrogenated forms of anions have been studied to understand any possible
contribution towards the observed AIBs.~Peripheral H atoms in a PAH anion molecule (C$_{24}$H$_{12}^{-1}$) have been
removed one$-$by$-$one to understand the effect of dehydrogenation on the spectra as shown in  Figure~\ref{fig2}. Usually,
H atoms are removed in a consecutive order, but in absence of a stable configuration, any preferable site for dehydrogenation
is considered provided that the energy is minimized.
As mentioned earlier, unlike neutrals all anions show rich 6$-$9~$\mu \rm m$ spectra which is characteristic of C$-$C 
stretch modes on the shorter wavelength side and C$-$H in$-$plane bending modes on the longer wavelength side.
The intermediate is a combination of the modes. As H atoms are removed, the C$-$C modes
gradually become dominant compared to the lesser number of C$-$H in$-$plane modes and eventually for a fully dehydrogenated
molecule (N$\rm_{H}$=~0)\footnote{N$\rm_{H}$=number of H atoms in the periphery of a PAH molecule}, C$-$C modes
are the only possible modes distributed in the 5$-$10~$\mu \rm m$ region. 
In fact, for (C$_{24}$H$_{n}$)$^{-1}$ with n=~8, 6, 4, 2, 1 \& 0,
the C$-$C mode is the most intense band (Int$\rm_{rel}\sim$~1)\footnote{Int$\rm_{rel}$=$\frac{\rm{absolute~intensity}}{\rm{maximum~absolute~intensity}}$} 
seen near 5.2 \& 6.8 $\mu \rm m$. The 5.2~$\mu \rm m$ band is a unique feature
observed for dehydrogenated PAHs only and arises due to the stretching of those C$-$C bonds that are
exposed on removal of its associated H atoms. The 5.2~$\mu \rm m$ band, however, does not appear 
immediately after dehydrogenation starts and arises only after certain dehydrogenation level is attained. 
In contrary to the predominance of the C$-$C stretch modes (5$-$10~$\mu \rm m$), a distinct 
fall in the intensity of the 3.3~$\mu \rm m$ is found. This is obvious as the 3.3~$\mu \rm m$ feature is purely a contribution from the C$-$H 
stretch mode. For (C$_{24}$)$^{-1}$, this feature is absent as expected. Besides this, the intensity of the 3.3~$\mu \rm m$ band is also affected
by the symmetry of the molecule that usually is less for a more symmetric structure.
When an odd number of H atoms are removed from the structure, the coronene molecule 
loses its symmetry by a degree lower compared to the removal of an even number of H atoms. 
A lower symmetry structure shows a higher intensity of the 3.3~$\mu \rm m$ feature
compared to a more symmetric structure.

Usually, a coronene molecule contains duo C$-$H groups~\footnote{A duo C$-$H group, 
also referred as doubly-adjacent C$-$H unit, is a group with one neighbouring adjacent C$-$H units 
on the same ring} at all sites. Removal of an odd number of H atoms from the periphery
leaves one of the associated C$-$H units to convert into a solo C$-$H 
unit~\footnote{A solo C$-$H group, also referred as non-adjacent C$-$H unit, is a group with no neighbouring adjacent C-H units 
on the same ring} which originally was a duo C$-$H group.
The stretching of such a converted solo C$-$H 
produces a 3.4~$\mu \rm m$ feature. Int$\rm_{rel}$ values for this feature are 0.95, 0.99,
1, 0.72, 1 and 0.45, respectively, for C$_{24}$H$_{11}^-$, C$_{24}$H$_{9}^-$, C$_{24}$H$_{7}^-$, C$_{24}$H$_{5}^-$, C$_{24}$H$_{3}^-$ and C$_{24}$H$_{1}^-$.
This is surprising as the molecules that originally have a solo C$-$H group
do not exhibit a 3.4~$\mu \rm m$ band. 
The astronomical 3.4~$\mu \rm m$ feature is proposed to 
be originated from PAH molecules with aliphatic or superhydrogenated side groups due to the aliphatic C$-$H stretch \citep[and references therein]{Yang17a}.

For coronene molecules containing solo C$-$H groups after dehydrogenation, 
an oop bending mode of the solo group is present at $\sim$~11.5~$\mu \rm m$.
We suspect that these significant shifts in the bands for a non-adjacent solo 
C$-$H unit arise due to the local electronic environment as the negative charge 
tends to be localised at the adjacent dehydrogenated site. 
As for the duo C$-$H groups of dehydrogenated molecules, the oop mode 
is redshifted and appears at $\sim$~12-13~$\mu \rm m$. The longer wavelength features beyond 13~$\mu \rm m$ are
mostly due to the C$-$C$-$C oop and C$-$C$-$C in$-$plane bending modes with a possible contribution from C$-$H oop modes
in a PAH molecule. Anionic coronene (C$_{24}$H$_{12}^{-1}$) shows features in the $~20-30~\mu \rm m$ region 
mostly due to the C$-$C$-$C in$-$plane modes which 
become significant as the dehydrogenation increases. These theoretically obtained features are unique and do
not appear for any other molecule studied so far.

We also performed similar calculation for a comparatively smaller PAH$-$ Perylene (C$_{20}$H$_{12}$) and
a comparatively larger PAH$-$ Ovalene (C$_{32}$H$_{14}$) to investigate any possible spectral variations. Figure \ref{fig3}
shows the theoretically obtained IR spectra for a few representatives of dehydrogenated forms 
of C$_{20}$H$_{12}^{-1}$ and C$_{32}$H$_{14}^{-1}$. 
The full table of the band positions and intensities of the modes for all the possible forms is 
available as supplementary material to the online version of this paper.
\subsection*{Perylene}
The intensity of the 3.3~$\mu \rm m$ feature decreases with the consecutive dehydrogenation similar
to coronene. The C$-$C stretch modes in the 5$-$10$~\mu \rm m$ region are present in the spectra of dehydrogenated perylene anions 
as can be seen from Figure \ref{fig3}(a$-$f). While this is expected for all forms of ionized PAHs, 
a unique features at $\sim$~5.8 is observed for 
dehydrogenated forms of perylene. This is analogous to the 5.2$~\mu \rm m$ band observed for a dehydrogenated coronene anion.
When H atoms are removed from a perylene molecule successively,
at a certain instance, one of the benzene rings in the network (apparently C$-$C$-$C) are completely free of H or any other atoms. 
Due to the stretching of that free C$-$C$-$C bond in dehydrogenated perylene, a feature at $\sim$~5.8~$\mu \rm m$ arises. The wavelengths 
beyond 10~$\mu \rm m$ are assigned to the  C$-$H oop and C$-$C$-$C oop modes, while features beyond 20~$\mu \rm m$ arise due to the 
C$-$C$-$C in$-$plane modes.

\subsection*{Ovalene}
\begin{figure*}
\hspace*{-5em}
\vspace*{-2em}
\includegraphics[width=15cm, height=18cm]{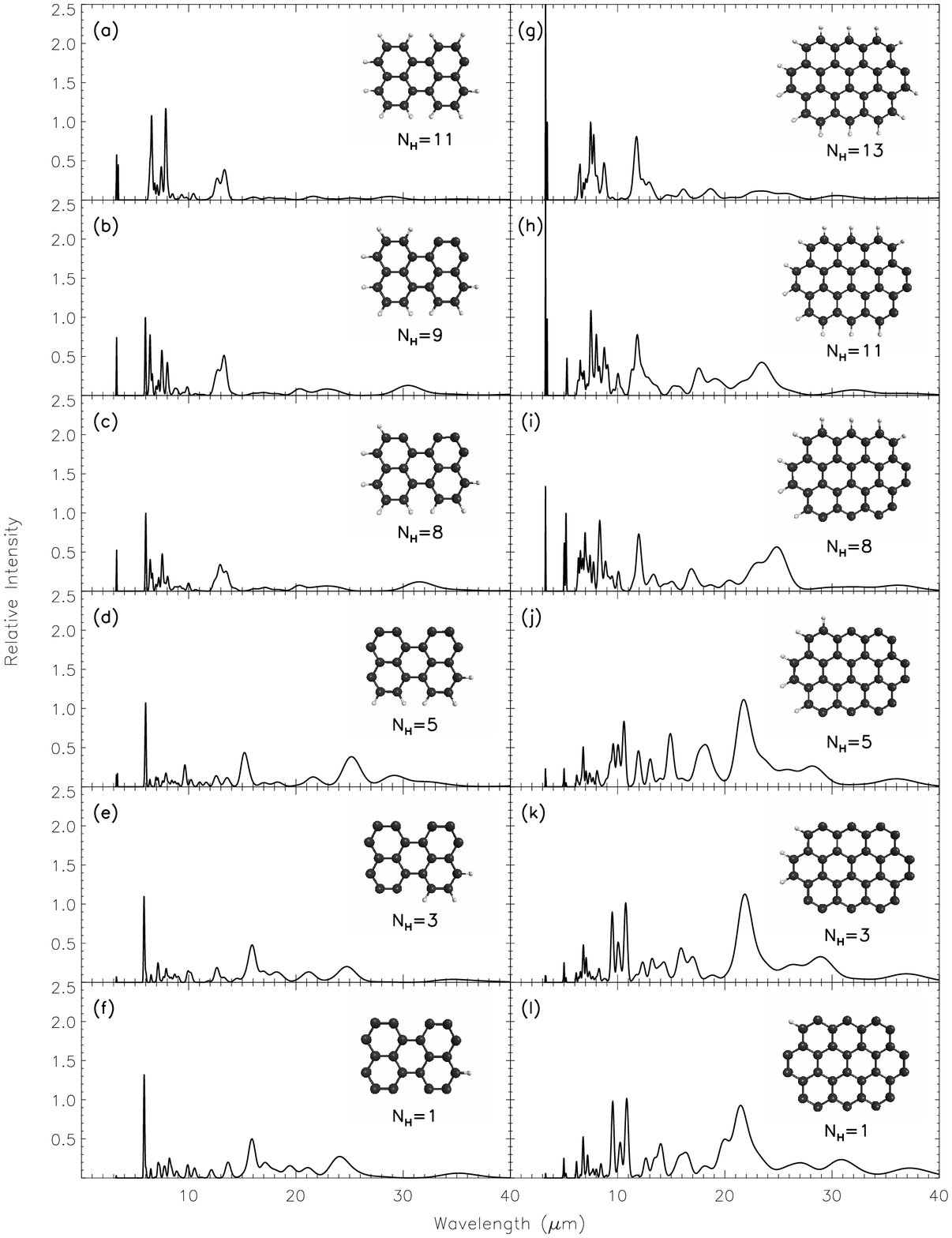}
\caption{Theoretical spectra of dehydrogenated forms of (a)-(f) perylene anion (C$_{20}$H$_{12}^{-1}$), (g)-(l) ovalene anion (C$_{32}$H$_{14}^{-1}$)}
\label{fig3} 
\vspace*{-0.5em}
\centering{
\raggedright N$\rm_{H}$=number of H atoms in the periphery of a PAH molecule}
\end{figure*}

Figure \ref{fig3}(g$-$l) show spectra of various forms of the dehydrogenated ovalene anion. 
With increasing dehydrogenation, the intensity of the 3.3~$\mu \rm m$ band decreases.
Apart from the other usual features that are common to all PAHs, dehydrogenated ovalene anions
show a 5.2$~\mu \rm m$ band due to a free C$-$C bond stretch. This is 
similar to the 5.2$~\mu \rm m$ feature observed for a dehydrogenated coronene and does not appear for a low degree of dehydrogenation.
Another crucial aspect is the increasing number of features with increasing size. 
Partial dehydrogenation of a comparatively large PAH anion (C$_{32}$H$_{14}^{-1}$) shows more  
features distributed thoroughout the $~5-30~\mu \rm m$ region compared to smaller dehydrogenated PAH anions.
Table~\ref{tab2} lists the wavelengths together with Int$\rm_{rel}$ values that are present in a partially dehydrogenated (50\% dehydrogenation) perylene anion (C$_{20}$H$_{6}^{-1}$),
coronene anion (C$_{24}$H$_{6}^{-1}$) and ovalene anion (C$_{32}$H$_{7}^{-1}$) that shows the enhancement of features in number with size.
\begin{table*}
\centering
  \begin{minipage}{90mm}
\caption{spectral lines in the 5$-$30~$\mu \rm m$ region in C$_{20}$H$_{6}^{-1}$, C$_{24}$H$_{6}^{-1}$ \& C$_{32}$H$_{7}^{-1}$}
\label{tab2}
\begin{tabular}[c]{c|c|cc}
\hline
\hline
  C$_{20}$H$_{6}^{-1}$  & C$_{24}$H$_{6}^{-1}$ & \ \ \ \ \ \ \ \ \ \ \ C$_{32}$H$_{7}^{-1}$   \\
 $\lambda$ ($\mu \rm m$)\ \ \ \ \ Int$\rm_{rel}$ & $\lambda$ ($\mu \rm m$)\ \ \ \ \ Int$\rm_{rel}$ & $\lambda$ ($\mu \rm m$)\ \ \ \ \ Int$\rm_{rel}$& $\lambda$ ($\mu \rm m$)\ \ \ \ \ Int$\rm_{rel}$\\ \hline
5.9 \ \ \ \ \ 1 & 5.1 \ \ \ \ \ 0.21 & \ 5 \ \ \ \ \ \ \ 0.56 & 9.1 \ \ \ \ \ 0.12 \\
\ \ \ 6.3 \ \ \ \ \ 0.07 & 5.2  \ \ \ \ \ \ \ \ 1 & 5.1 \ \ \ \ \ 0.96 &  9.3 \ \ \ \ \ 0.17 \\
\ \ \ 6.3 \ \ \ \ \ 0.19 & 6.9 \ \ \ \ \ 0.28 & 5.2 \ \ \ \ \ \ \ \ \ \  1 &  9.5 \ \ \ \ \ 0.11 \\
\ \ \ 6.4  \ \ \ \ \ 0.60 & 7.1 \ \ \ \ \ 0.07 & 6.2 \ \ \ \ \ 0.05 &   9.7 \ \ \ \ \ 0.09 \\
\ \ \ 6.7 \ \ \ \ \ 0.06 & 7.3 \ \ \ \ \ 0.21 & 6.3 \ \ \ \ \ 0.22 & 10.1 \ \ \ \ \ 0.23 \\ 
\ \ \ 7.2 \ \ \ \ \ 0.18 & 7.5 \ \ \ \ \ 0.08 & 6.3 \ \ \ \ \ \ \ 0.1 & 11.6 \ \ \ \ \ 0.05 \\
\ \ \ 7.2  \ \ \ \ \ 0.11 & 8 \ \ \ \ \ \ \ 0.09 & 6.5 \ \ \ \ \ 0.19 & 11.6 \ \ \ \ \ 0.08 \\
\ \ \ 7.4  \ \ \ \ \ 0.21 & 8 \ \ \ \ \ \ \ 0.05 &  6.6 \ \ \ \ \ 0.46 &  11.9 \ \ \ \ \ \ \ 0.6 \\
\ \ 7.5 \ \ \ \ \ 0.3 & 9.1 \ \ \ \ \ 0.11 & 6.7 \ \ \ \ \ 0.05 & 12.3 \ \ \ \ \ 0.16 \\ 
\ \ \ 7.5  \ \ \ \ \ 0.14  & 9.4 \ \ \ \ \ 0.13 &6.8 \ \ \ \ \ 0.09 &  12.3 \ \ \ \ \ 0.06 \\
\ \ \ 7.9 \ \ \ \ \  0.10 & 10.1 \ \ \ \ \ 0.05 &  6.8 \ \ \ \ \ 0.75 & 12.8 \ \ \ \ \ 0.15 \\ 
\ \ \ 8.5  \ \ \ \ \ 0.17 & 10.7 \ \ \ \ \ 0.07 & 6.9 \ \ \ \ \ 0.24 &  13.3 \ \ \ \ \ 0.05 \\
\ \ \ 8.7  \ \ \ \ \ 0.14 & 12.2 \ \ \ \ \ 0.22 & 6.9 \ \ \ \ \ 0.73 &  13.3 \ \ \ \ \ 0.08 \\
\ \ \ 8.9  \ \ \ \ \ 0.21 & 12.6 \ \ \ \ \ 0.06 &  7.1 \ \ \ \ \ 0.25 & 14.5 \ \ \ \ \ 0.16 \\
\ \ \ 9.9  \ \ \ \ \ 0.28 & 12.7 \ \ \ \ \ 0.05 & 7.1 \ \ \ \ \ 0.16 &  14.6 \ \ \ \ \ \ \ 0.1  \\
\ \ \ 9.9 \ \ \ \ \ 0.06 &  16.3 \ \ \ \ \ 0.09 &  7.3 \ \ \ \ \ 0.12 &  14.6 \ \ \ \ \ 0.06 \\
\ \ 12.4 \ \ \ \ \ 0.08 & 19.5 \ \ \ \ \ 0.06 &  7.3 \ \ \ \ \ 0.13 & 15.1 \ \ \ \ \ 0.31 \\ 
\ \ 12.5  \ \ \ \ \ 0.25 & 20.1 \ \ \ \ \ 0.05 & 7.7 \ \ \ \ \ 0.13 &  16.7 \ \ \ \ \ \ \ 0.3 \\
\ \ 13.2 \ \ \ \ \  0.35 & 22.1 \ \ \ \ \ 0.09 & 7.8 \ \ \ \ \ \ \ 0.1  &  16.9 \ \ \ \ \ \ \ 0.1 \\
\ \ 16.6  \ \ \ \ \ 0.09 & 23.4\ \ \ \ \ 0.28 & 7.8 \ \ \ \ \ 0.05 & 17.1 \ \ \ \ \ 0.07  \\
\ \ 26.9  \ \ \ \ \ 0.09 & 26.8\ \ \ \ \ 0.06 & 7.9 \ \ \ \ \ 0.07 & 18.2 \ \ \ \ \ \ \ 0.2 \\  
\ \ 28.4  \ \ \ \ \ 0.08 & 30.3  \ \ \ \ 0.08 & 8.2 \ \ \ \ 0.07 & 19.9 \ \ \ \ \ 0.09 \\ 
   \ \ \ \ \  & & 8.2 \ \ \ \ \ 0.82 &  20.3 \ \ \ \ \ 0.05 \\ 
  \ \ \ \ \  & & 8.4 \ \ \ \ \ 0.24 & 22.6 \ \ \  \ \ 0.21 \\  
     \ \ \ \ \  & &  8.5 \ \ \ \ \ 0.07 &   22.7 \ \ \ \ \ 0.54 \\  
     \ \ \ \ \  & & 8.7 \ \ \ \ \ 0.05 &  24.1 \ \ \ \ \ 0.09 \\ 
      \ \ \ \ \  & & 8.9 \ \ \ \ \ 0.21 &  24.4 \ \ \ \ \ 0.09 \\  
       \ \ \ \ \  & & 8.9 \ \ \ \ \ 0.07  &  25.0 \ \ \ \ \ 0.46  \\  
   \ \ \ \ \  & &  &  26.0 \ \ \ \ \ 0.19  \\
     \hline
\end{tabular}
The full table of the band positions and intensities of the spectral lines for all the molecules is 
available as supplementary material to the online version of this paper.
\end{minipage}
\end{table*}

\section*{Astrophysical implications}
Recent studies by \citet[]{Bauschlicher08, Bauschlicher09} conclude that PAH neutrals, cations and anions are
equally important for some of the observed AIBs, if not all. 
Particularly, the 3.3~$\mu \rm m$ feature is attributed to arise 
from the C$-$H stretch mode in a neutral PAH whereas the 6.2, 7.7 and 8.6~$\mu \rm m$ features are inherent to 
C$-$C stretch/C$-$H in$-$plane modes in a PAH cation \citep[]{Pech02}. The 11.2 and 12.7~$\mu \rm m$ features have been attributed
to the oop bending modes associated with solo and duo/trio C$-$H groups respectively in a neutral PAH \citep[]{Hudgins99}. 
In general, PAH cations are proposed to be promising candidates of AIB carriers in terms of intensities, while neutral PAHs are consistent
with the band profiles and peak positions of the bands \citep[and references therein]{Pech02}.
The ionization (positive/negative) will result in a slight blueshifting of the peak position with a distinct variation in the intensity. 

Anionic PAHs have been considered to be the principal carriers of negative charge in diffuse and dense interstellar clouds 
\citep[]{Bakes&Tielens98, Wakelam08, Tielens08, Carelli12}. Similar to cations, PAH anions may also contribute to the astronomical bands 
at 7.8~$\mu \rm m$ and 8.6~$\mu \rm m$ \citep[]{Bauschlicher08, Bauschlicher09}. The variation in the intensity of the IR bands
is a function of the physical condition of the environment and can be used to probe the abundance of PAH cations/anions in the sources.
Our results suggest that PAH anions show spectral characteristics that are in-between PAH neutrals and PAH cations.
We also include  dehydrogenated forms of PAH anions to understand any possible contribution towards the observed AIBs.
PAH anions are formed through radiative attachment of electrons in dense cloud. There is a possible channel for stabilization of PAH anion by
releasing a H atom to form dehydrogenated PAH anion \citep[]{Sebastianelli12, Garcia13}. The higher electron affinity
of dehydrogenated PAH molecules in comparison to the parent hydrocarbon also indicates 
the possibility of formation of dehydrogenated PAH anions in the ISM.
\citet[]{LePage01, LePage03} reported that intermediate size PAHs (carbon atoms $\sim$~20 to 30) are prone to extreme dehydrogenation, while
small PAHs (fewer than 20 carbon atoms) are destroyed and large PAHs (more than 30 carbon atoms) are mostly hydrogenated/fully hydrogenated 
under interstellar conditions. Those with dehydrogenation can further be altered into a
dehydrogenated PAH anion in dense clouds.
Similar to neutral PAHs, anionic PAHs also show a distinct
3.3~$\mu \rm m$ band with a comparatively stronger intensity. 
\begin{table*}
 \centering
  \begin{minipage}{80mm}
 \caption{Int$_{3.3}$ obtained for dehydrogenated forms of coronene anion}
 \label{tab3}
 \begin{tabular}[c]{c|c}
 \hline \hline
dehydrogenated coronene anion & Int$_{3.3}$ \\ \hline
C$_{24}$H$_{11}^{-1}$ &  0.63 \\ \hline
C$_{24}$H$_{10}^{-1}$  & 0.92 \\ \hline
C$_{24}$H$_{9}^{-1}$  &  0.99  \\ \hline
C$_{24}$H$_{8}^{-1}$  & 0.37 \\ \hline
C$_{24}$H$_{7}^{-1}$ & 1 \\ \hline
C$_{24}$H$_{6}^{-1}$   &  0.16 \\ \hline
C$_{24}$H$_{5}^{-1}$   &  0.72 \\ \hline
C$_{24}$H$_{4}^{-1}$  &  0.21 \\ \hline
C$_{24}$H$_{3}^{-1}$  &  1 \\ \hline
C$_{24}$H$_{2}^{-1}$  &  0.07 \\ \hline
C$_{24}$H$_{1}^{-1}$  &  0.45 \\ \hline
\end{tabular}
\end{minipage}
 \end{table*}
From this report, Int$_{3.3}$\footnote{The highest C$-$H$_{\rm{stretch}}$ intensity among all the C$-$H$_{\rm{stretch}}$ vibrations is considered} 
for neutral coronene (C$_{24}$H$_{12}$) is 0.44 and 
for anionic coronene (C$_{24}$H$_{12}^{-1}$), it is 0.59.
Int$_{3.3}$ is large compared to the observations.
\citet[]{Tielens08} pointed out that neutral PAHs can still be considered as potential candidates for the 3.3~$\mu \rm m$
band as long as the relative strength of the C$-$H versus the C$-$C modes is comparable to observations. 
However for the anions, Int$_{3.3}$ is so large that it only becomes comparable to observations when 
H atoms are removed from the periphery of the molecule.
Dehydrogenation may not necessarily result into an immediate decrease in Int$_{3.3}$ as the intensity is also 
governed by the symmetry of the molecule. PAHs with higher symmetry show a lower value of Int$_{3.3}$ compared
to less symmetric structures. 
As dehydrogenation proceeds, we expect to see
a fall in Int$_{3.3}$ and after a certain instance, it may give an Int$_{3.3}$ which is equivalent to that of neutral and gradually may approach 
the Int$_{3.3}$ value obtained for a PAH cation. For example, C$_{24}$H$_{8}^{-1}$ has 
Int$_{3.3}\sim$~0.37 which is then close to the value obtained for C$_{24}$H$_{12}$. Thus, C$_{24}$H$_{8}^{-1}$ may be equally important as
C$_{24}$H$_{12}$ for the astronomically observed feature at 3.3~$\mu \rm m$. The low symmetry structures 
of dehydrogenated coronene containing an odd number of H atoms usually
show a higher value of Int$_{3.3}$ compared to high symmetric structures
with an even number of H atoms as shown in Table~\ref{tab3}. C$_{24}$H$_{2}^{-1}$
gives Int$_{3.3}\sim$~0.07 which is close to the Int$_{3.3}\sim$~0.02 obtained for C$_{24}$H$_{12}$ cation.
Perylene (C$_{20}$H$_{12}$) also shows similar behaviour and on partial dehydrogenation, C$_{20}$H$_{10}^{-1}$ gives Int$_{3.3}$ $\sim$~0.59 which is
close to the Int$_{3.3}$ obtained for C$_{20}$H$_{12}$ (Int$_{3.3}$ $\sim$~0.69).
Int$_{3.3}$ for the C$_{20}$H$_{12}$ anion is high ($\sim$~0.93).
Increasing dehydrogenation will give even a lower value of Int$_{3.3}$ as discussed for coronene.
As for ovalene (C$_{32}$H$_{14}$), the situation is somewhat different and Int$_{3.3}$ obtained for its neutral, anionic and 
cationic forms are 0.83, 0.29, 0.07 respectively.
Int$_{3.3}$ increases to 1 when we remove the first H atom from C$_{32}$H$_{14}^{-1}$ (i.e. for C$_{32}$H$_{13}^{-1}$) and it then comes down to a
lower value for a C$_{32}$H$_{10}^{-1}$ (Int$_{3.3}$ $\sim$~0.37), C$_{32}$H$_{6}^{-1}$ (Int$_{3.3}$ $\sim$~0.32) and so on.
For C$_{32}$H$_{3}^{-1}$ and C$_{32}$H$_{1}^{-1}$, Int$_{3.3}$ is 0.08 and 0.06 respectively which are comparable to observations. 

Dehydrogenation also results in the rise of features in the 
5$-$10~$\mu \rm m$ region. The same characteristic is also observed for cationic PAH molecules.
A band at 5.2~$\mu \rm m$ is an unique feature present only in strongly dehydrogenated PAH anions. This feature may help
in identification of dehydrogenated PAH anions in the ISM. Its absence also suggests that the size of PAHs studied here may be only weakly dehydrogenated.
Another important characteristic is that with increasing dehydrogenation, 
a broad plateau in the 20$-$30~$\mu m$ region gains intensity mainly due to C$-$C$-$C in$-$plane modes. With increasing size,
this feature gains intensity. This region together with the other characteristics mentioned here may be used as a probe to identify any possible dehydrogenated
forms of PAH anions in the ISM. [FePAH]$^+$ complexes also show intense features in the 20$-$60~$\mu \rm m$ regions \citep[]{Simon10}, however the assignment
of carriers in the far-infrared region is still ambiguous. This calls for new astronomical data in the far-infrared which would help concretize theoretical results. 

It is important to note here that interstellar PAHs usually exist either in fully hydrogenated
or fully dehydrogenated form and the intermediate hydrogenated/dehydrogenated state is rare \citep[]{Mackie15}. 
\citet[]{LePage01, LePage03} suggest that partial dehydrogenation may be 
possible for intermediate size PAHs (carbon atoms $\sim$ 20 to 30).
In such a scenario, intermediate dehydrogenation states for C$_{32}$H$_{14}$ may exist long enough to emit the AIBs. 
Once all the H atoms are removed, a C atom may be lost subsequently to form a pentagonal ring which
will then initiate the formation of fullerenes in the ISM \citep[]{Mackie15}. Dehydrogenation may therefore be considered as an intermediate phase in the 
formation of a fullerene from a PAH molecule.

\section*{Summary}
Interstellar PAH anions are likely to exist in cold dense regions where a PAH molecule can easily attract an electron to form a PAH anion
that shows important characteristic features in IR. This study reports the vibrational study of dehydrogenated PAH anions
to seek any correlation with the observed AIBs. The theoretical data shows that PAH anions along with their dehydrogenated
forms show characteristics that are similar to neutrals and cations. This implies that the astronomically
observed bands that are usually assigned to either a neutral or a cationic
PAH are also likely to have contributions from dehydrogenated PAH anions
provided that the physical conditions support their presence in the ISM. 

We summarize the results here:
\begin{description}
 \item[1.] Partially dehydrogenated PAH anions may contribute to the 3.3~$\mu \rm m$ region and the broad intense features in the 5$-$10 $\mu \rm m$ region.
\item[2.] The 3.3~$\mu \rm m$ feature is immensely strong for a PAH anion. Dehydrogenation of a PAH anion
reduces the intensity of the 3.3~$\mu \rm m$ feature which then becomes comparable 
to its neutral counterpart and is closer to observed intensities of regions where the 3.3~$\mu \rm m$ is intense. 
\item[3.] During the process of successive dehydrogenation, the molecule passes through a series of changes in its symmetry. The dehydrogenated PAH anion
with a higher symmetry produces a less intense 3.3~$\mu \rm m$ band than that with a lower symmetry.
\item[4.] During the process of dehydrogenation, a duo C$-$H group in the PAH anion may convert into a solo C$-$H unit which produces a 3.4~$\mu \rm m$ feature
due to the solo C$-$H stretch vibrations.
\item[5.] Unique features at 5.2 and 5.8~$\mu \rm m$ are observed for a dehydrogenated PAH anion 
that arise due to the stretching of the free C$-$C/C$-$C$-$C stretch. The 5.2~$\mu \rm m$ band
observed in coronene and ovalene may be used to identify dehydrogenated PAH anions. Absence of this feature
in observations suggests that such PAHs may be dehydrogenated up to a certain limit.
\item[6.] A broad plateau in the 20$-$30~$\mu \rm m$ region arises in the spectra of a dehydrogenated PAH anion that becomes 
more significant with increasing size.
\end{description}

These characteristics can be used as a tool to understand any possible contribution of dehydrogenated PAH anions in the ISM. 
The calculated energy configuration indicates that before they convert into fully dehydrogenated molecules, a partially dehydrogenated form of PAH anion
may be present for a large PAH molecule as it is less susceptible to the loss of H atoms and requires time to remove all the H atoms from its periphery. 
The spectral characteristics together with stability offers support to the existence of large dehydrogenated PAH anions 
in the ISM as members of the extended PAH family. 
\section*{Acknowledgements}
AP acknowledges financial support from ISRO Respond grant (ISRO/RES/2/401/15-16) and DST EMR grant, 2017. 
AP thanks the Inter-University Centre for Astronomy and Astrophysics, Pune for associateship. 
PJS thanks the Leverhulme Trust for the award of a Leverhulme Emeritus Fellowship. 
\def\aj{AJ}%
\def\actaa{Acta Astron.}%
\def\araa{ARA\&A}%
\def\apj{ApJ}%
\def\apjl{ApJ}%
\def\apjs{ApJS}%
\def\ao{Appl.~Opt.}%
\def\apss{Ap\&SS}%
\def\aap{A\&A}%
\def\aapr{A\&A~Rev.}%
\def\aaps{A\&AS}%
\def\azh{AZh}%
\def\baas{BAAS}%
\def\bac{Bull. astr. Inst. Czechosl.}%
\def\caa{Chinese Astron. Astrophys.}%
\def\cjaa{Chinese J. Astron. Astrophys.}%
\def\icarus{Icarus}%
\def\jcap{J. Cosmology Astropart. Phys.}%
\def\jrasc{JRASC}%
\def\mnras{MNRAS}%
\def\memras{MmRAS}%
\def\na{New A}%
\def\nar{New A Rev.}%
\def\pasa{PASA}%
\def\pra{Phys.~Rev.~A}%
\def\prb{Phys.~Rev.~B}%
\def\prc{Phys.~Rev.~C}%
\def\prd{Phys.~Rev.~D}%
\def\pre{Phys.~Rev.~E}%
\def\prl{Phys.~Rev.~Lett.}%
\def\pasp{PASP}%
\def\pasj{PASJ}%
\def\qjras{QJRAS}%
\def\rmxaa{Rev. Mexicana Astron. Astrofis.}%
\def\skytel{S\&T}%
\def\solphys{Sol.~Phys.}%
\def\sovast{Soviet~Ast.}%
\def\ssr{Space~Sci.~Rev.}%
\def\zap{ZAp}%
\def\nat{Nature}%
\def\iaucirc{IAU~Circ.}%
\def\aplett{Astrophys.~Lett.}%
\def\apspr{Astrophys.~Space~Phys.~Res.}%
\def\bain{Bull.~Astron.~Inst.~Netherlands}%
\def\fcp{Fund.~Cosmic~Phys.}%
\def\gca{Geochim.~Cosmochim.~Acta}%
\def\grl{Geophys.~Res.~Lett.}%
\def\jcp{J.~Chem.~Phys.}%
\def\jgr{J.~Geophys.~Res.}%
\def\jqsrt{J.~Quant.~Spec.~Radiat.~Transf.}%
\def\memsai{Mem.~Soc.~Astron.~Italiana}%
\def\nphysa{Nucl.~Phys.~A}%
\def\physrep{Phys.~Rep.}%
\def\physscr{Phys.~Scr}%
\def\planss{Planet.~Space~Sci.}%
\def\procspie{Proc.~SPIE}%
\let\astap=\aap
\let\apjlett=\apjl
\let\apjsupp=\apjs
\let\applopt=\ao

\bibliographystyle{mn} 
\bibliography{mridu}

\end{document}